\documentclass{elsart}

\usepackage{graphicx}
\usepackage[]{epsfig}
\usepackage{xspace}
\usepackage{amssymb} 
\usepackage{psfrag}
\usepackage[ansinew]{inputenc}
\usepackage{newlfont}

\newcommand{\wimp}{WIMP\xspace}
\newcommand{\wimps}{WIMPs\xspace}

\newcommand{\ca}{CaWO$_4$\xspace}
\newcommand{\fe}{$^{55}$Fe\xspace}
\newcommand{\MPI}{Max-Planck-Institut\xspace}
\newcommand{\cresst}{CRESST\xspace}

\newcommand{\gam}{$\gamma$\xspace}

\newlength{\myVSpace}
\newcommand\xstrut{\raisebox{-.2\myVSpace}
  {\rule{0mm}{\myVSpace}}%
}
\newlength{\maliVSpace}
\begin{document}

\begin{frontmatter}

\title{New Technique for the Measurement of the Scintillation Efficiency of Nuclear Recoils}

\author{Jelena Ninkovi\'{c}\corauthref{cor1}},
\ead{ninkovic@mppmu.mpg.de}
\corauth[cor1]{Corresponding author: tel: +49 89 83940049; fax: +49 89 83940011;}
\author{Peter Christ,}
\author{Godehard Angloher,}
\author{Dieter Hauff,}
\author{Partick Huff},
\author{Emilija Panti\'{c},}
\author{Federica Petricca,}
\author{Franz Pr\"{o}bst,}
\author{Wolfgang Seidel }
\address{\MPI f\"{u}r Physik (Werner-Heisenberg-Institut)\\F\"{o}hringer Ring 6, D-80805 Munich, Germany}
\journal{NIM A}

\begin{abstract}
We present a new technique developed for the measurement of the scintillation efficiency of nuclear recoils in solid scintillators. Using this technique we measured the quenching of the scintillation efficiency for the various recoiling nuclei in \ca crystals which are used in direct Dark Matter searches.
\end{abstract}

\end{frontmatter}

\section*{Introduction}
\parindent 2em
The direct detection and the understanding of the nature of Dark Matter remain the outstanding challenges of present day physics and cosmology. Searches for Weakly Interacting Massive Particles (WIMPs) which may constitute the galactic Dark Matter are presently being carried out by a number of groups world-wide \cite{cdms2004}\cite{edelweiss2005}\cite{cresst2}. Direct detection experiments appear today as one of the most promising techniques to detect particle Dark Matter. All direct detection experiments rely on the basic idea that many \wimps should pass through the Earth, making their detection possible. Interaction with ordinary matter is expected via elastic scattering on nuclei which, for the spin-independent case, should be coherent with all nucleons in a nucleus resulting in a cross-section rising in proportion to $A^2$ (where $A$ is the mass number) and therefore favoring heavy nuclei \cite{kamionkowski}. In addition a low energy threshold is an experimental requirement as the predicted energy spectrum falls exponentially with energy. Due to the extremely low \wimp\,-\,nucleus scattering cross section the ability to discriminate \wimp signals from the radioactive background is essential for this kind of experiments.

Cryogenic detectors based on simultaneous detection of light and heat fulfill these requirements \cite{meunier}. They utilize the fact that the light output from nuclear recoil events in the scintillator is reduced compared to the light observed from $\beta$ and $\gamma$\,--\,interactions (``electron recoils'') of the same energy. This is characterized by the $\underline{Q}$uenching $\underline{F}$actor (QF) which is defined as the ratio of the light output from electron recoils and from nuclear recoils of the same energy. For an effective background suppression the quenching factors have to be known, especially because different nuclei present in the scintillator are  expected to have different quenching factors \cite{Lin65}\cite{lindhard}. In contrast to the \wimps, the residual neutron background comes mostly from elastic scattering off the lighter elements. Therefore the different quenching of different recoiling nuclei in the scintillator can be used for a discrimination of the neutron background.

In order to measure the quenching factors of different recoiling nuclei in scintillators we have developed a new technique. It relies on the measurement of the light output produced when the scintillating crystal is bombarded by different ions. This method allows to measure quenching factors for a wide range of different elements and scintillators.

\section{Experimental technique}
Nuclear recoil quenching factors have been measured for a variety of scintillators such as NaI \cite{Spooner1994}\cite{tovey1998}, CaF$_2$(Eu) \cite{tovey}, CsI(Tl) \cite{horn} with neutron scattering
experiments.

Instead of measuring the light produced by the recoiling nuclei of the target material as in the neutron scattering experiments, the technique presented here measures the light emitted when single ions impinge onto the scintillating crystal. As this work was done in the framework of the \cresst\,\footnote{\underline{C}ryogenic \underline{R}are \underline{E}vent \underline{S}earch with \underline{S}uperconducting \underline{T}hermometers} experiment, \ca was used as target material. Ions with fixed kinetic energy were produced with a Laser Desorption\,/\,Ionization ion source (LDI) in a time-of-flight mass spectrometer (TOF-MS). The \ca crystal was mounted at the end of the flight tube and the light was detected using a photomultiplier. An \fe source was used for the reference measurement.

\subsection{TOF mass spectrometer with an LDI ion source}\label{sec:mass spec}
Figure 1 shows the scheme of the time-of-flight mass spectrometer with the LDI ion source used for this measurement. The ion source was separated with a small gate valve from the main vacuum chamber of the spectrometer. This allowed to vent just the small compartment for the exchange of the target, with minimal interruption ($\simeq$~20\,min) of operation.

Six different target materials could be fixed on the target holder, using silver epoxy (Fig.\,2). The target holder could be rotated and moved transversely to be able to reach the whole target with the laser.

A shot from the pulsed UV laser of 337\,nm wavelength desorbed and ionized the atoms. The shot duration was in the order of nanoseconds. The laser power could be controlled with a dielectric attenuator and the beam was focused on the sample disc with a lens of 200\,mm focal length. At the sample target the laser power density was typically 10$^6$\,--\,10$^7$\,W/cm$^2$. The maximum acceleration voltage of the LDI ion source was 18\,keV. The ion beam was focused on the scintillator using an Einzel lens and X--Y deflection plates (Fig.\,1).

The MS was instrumented with an additional ion deflector which allowed the selection of ions with a certain mass. It consisted of a pair of plates the electric field of which was turned on when the laser shot was fired. It deflected all ions out of the beam path until the electric field was switched off. Therefore, all lighter ions preceding the ions of interest were deflected. The high voltage was then switched off to allow the passage of the desired ions and then switched on again in order to block higher mass ions (see Fig.\,3).

The time-of-flight part used for the quenching factor measurements consisted of two ion reflectors (for details see \cite{proteom}). Between these two reflectors a micro channel plate (MCP) could be inserted into the ion path at any moment of operation (Fig.\,1). It was used to define the time window for switching off the voltage applied to the deflection plates to allow only the passage of the ion species of interest from a contaminated target material or from a multielement target (see Fig.\,3). The selected elements and target materials are listed in table\,\ref{tab:elements}. The time-of-flight measured with the MCP, which is roughly half of the one at the outlet of the mass spectrometer, was used to adjust the photon counting window appropriately for each element (see Sec. \ref{sec:QFroomtempmethod} and Fig.\,5).

The laser pulse was used as a start trigger and the signals of the MCP for the time-of-flight measurement were read out using a 500\,MHz PCI transient digitizer (Fast Com Tec - TRP 250/500).

\subsection{PMT setup for photon counting}\label{sec:QFroomtempmethod}
\noindent The photon counting setup consisted of a small \ca crystal (5$\times$5$\times$5\,mm$^3$) held in a teflon reflector and viewed by a photomultiplier tube that was placed in a housing which was mounted at a gate valve at the outlet of the mass spectrometer (Fig.\,1 and Fig.\,4). Ions reached the \ca crystal via a small hole (\o\,=\,1\,mm, length\,=\,8\,mm) in the reflector. In order to reduce the effect of trapped light in the crystal, the crystal surface facing the photomultiplier tube (PMT) was roughened\,\footnote{Polished with 10\,$\mu$m grain size.}. Additionally, a thin layer of silicon grease was placed between the crystal and the PMT for optical coupling. The PMT was an Electron Tubes 9124B selected for a low rate of dark counts. A preamplifier followed by a 500\,MHz digitizer was used for the read out of the PMT signal. The data acquisition was triggered by the laser. For each laser shot the number of photons was determined by counting the number of voltage pulses exceeding a given threshold amplitude in a fixed time window after the arrival of the ion type of interest (see Fig.\,5). The number of registered photons for each shot was then accumulated in a histogram. Additionally, for each registered photon its arrival time was collected in a second histogram, which gave the light curve of the scintillator and the time-of-flight of the ions (see Fig.\,7).

The reference measurement for the quenching factor was made with 5.9\,keV X\,-\,rays from an \fe source. The radioactive source could be installed in front of the collimator without any change of the setup. For the reference measurement the data taking was triggered by the PMT signal. The photon counting was performed as in the case of ions.

\section{Results and Discussion}

\subsection{MCP measurements}
The elements (H, O, Si, Ca, Cu, Y, Ag, Sm, W, Au) were chosen to smoothly cover the broad mass range of the elements in \ca (see table \ref{tab:elements}).
\newline
For a proper placement of the time window for photon counting the arrival time of each ion type was measured with the MCP. As shown in Fig.\,6\,(left) the arrival times measured with the MCP follow the expected linear dependence on $\sqrt{A}$\,\,\,\,\footnote{Neglecting the time-of-flight in the ion source, the flight time $t$ is
\begin{equation}\label{eq:tof}
    t\,=\,D\,\sqrt{\frac{1}{2eU}\,\frac{A}{z}}\,.
\end{equation}
where $A$ is the mass number of the ions, $U$ is an electric potential difference, $z$ charge state of the ions and $D$ is the flight distance. }.
\subsection{PMT measurements}
The width of the counting time window was kept fixed (30, 40, 50~$\mu s$) for all measured elements.
Figure\,7-left shows light curves of \ca obtained with beryllium and copper ions. The times of flight extracted from the onset of the light curve (Fig.\,6\,right) were roughly twice as long as the ones measured with the MCP, consistent with the longer flight distance.

The sharp rise time ($<50\,ns$) of the recorded light curves confirms the absence of electrical fields due to a charge up of the crystal or of its teflon holder during the measurement. Such charging up would slow down the ions before hitting the scintillator and therefore change their time of flight and this in turn would broaden the rise of the light curve or shift the onset position of the light curve in consecutive measurements. A comparison of several subsequently measured light curves confirmed the absence of such charging up.

Figure\,7-right shows the biexponential fit of the measured light curve. The faster decay time determined this way is 1.08\,$\mu s$ while the longer one is 7.96\,$\mu s$. Measured decay times are in the agreement with recently reported values \cite{Zde04}.

\subsubsection{Reference measurement with X-rays}
The electron recoil reference measurement for the determination of the quenching factors was made with 5.9\,keV X-rays from an \fe source. Figure\,8 shows the spectrum measured with a 40\,$\mu$s photon counting window. Simultaneous measurements with three time windows (30, 40, 50\,$\mu$s) were made in order to probe the dependence of the results on the width of the chosen time window. The spectra have been fitted using the fit procedure developed for ion spectra which will be discussed later. The results are presented in Table\,\ref{tab:reference}.

\subsubsection{Spectra measured with ions}
The laser power and the ion beam optics were adjusted such that the probability of multiple ion events (more than one ion impinging on the crystal surface per laser shot) was reduced as much as possible. The procedure is illustrated in figure\,9. Laser power and focus were tuned that on average only each 15th laser shot gave an ion signal (Fig.\,9\,bottom).

\subsubsection{Data analysis}
For each laser shot a certain number of photons was detected. The main contribution to the recorded spectra comes from laser shots not producing ion signals and thus measure the dark counts of the photomultiplier. The number of dark counts obeys a Poisson distribution with a small mean value. The number of the photons produced by single ion events\,\footnote{Due to the reduced laser power and applied beam defocusing only single and a smaller number of double ion events are occurring.} gives a spectrum with a Poisson distribution with a higher mean value. If multiple ion events exist additional Poissonians will appear with a mean value equal to a multiple of the mean of the first one. In the counting procedure the origin of the registered photons cannot be identified. Therefore, the background had to be accounted for in the fit function.
\newline
For simplicity first only none (background) and one ion (signal) arriving per laser shot will be considered. In this case the probability to have $k$ registered photons, $P_t(k)$, is given by
\begin{equation}\label{eq:simpleFit}
   P_t(k)\,=\,P_{ion}(0)\,\cdot\,P(k,\overline{B})\,+\,P_{ion}(1)
     \,\sum_{l=0}^{k}\,P(l,\overline{B}) \,P_S(k-l,\overline{n}\,\cdot\,1)\,,
\end{equation}
where $P_{ion}(i)$ is the probability for $i$ ions arriving per laser shot, $P(k,\overline{B})$ is the probability of observing $k$ background counts from a Poissonian distribution with mean $\overline{B}$ and  $P_S(m,\overline{n})$ is the probability that $m$ signal counts are observed when $\overline{n}$ is the average number of observed photons for single ion events. The term describing the ion signal sums the probabilities of all possible contributions of background and signal counts. When up to $N$ ions arrive per laser shot equation \ref{eq:simpleFit} generalizes to
\begin{equation}\label{eq:Fit}
      P_t(k)\,=\,P_{ion}(0)\,\cdot\,P(k,\overline{B})\,+\,\sum_{i=1}^{N}P_{ion}(i)\,
       \sum_{l=0}^{k}\,P(l,\overline{B}) \,P_S(k-l,\overline{n}\,\cdot\,i)\,.
\end{equation}
Here, $P_{ion}(i)$, $\overline{n}$ and $\overline{B}$ are the parameters to be determine by the experiment. The background is described by a Poissonian with a scale factor $P_{ion}(0)$. The mean value $\overline{B}$ of the background Poissonian is determined by a separate background measurement for each time window. The ion signal can be fitted with a selectable number of multiple ion events to be taken into consideration. The scale factor $P_{ion}(i)$ for each of the signal Poissonians i.e. the probability that a number $i$ of ions arrives per laser shot and the average number $\overline{n}$ of detected photoelectrons per ion are the remaining free fit parameters. In figure\,10 different fits to the photon multiplicity spectra measured for H and W ions are shown. The dashed lines represent fitted spectra using the method described above.

As the peak positions of the single and double ion events of the hydrogen spectra are well separated from the background it was possible to check the validity of the background model used in the fit. Therefore, the three scale factors ($P_{ion}(0)$, $P_{ion}(1)$, $P_{ion}(2)$), $\overline{n}$ and $\overline{B}$ were left as free fit parameters. Within the statistical errors the values obtained for $\overline{B}$ are in agreement with those from the separate background measurements (for the 30\,$\mu$s time widow determined $\overline{B}$ values are: 0.0180$\pm$0.8E-03 from the separate background measurement while 0.0176$\pm$0.12E-02 from the H spectrum.). To avoid the influence of the correlation between fit parameters $\overline{n}$ and $\overline{B}$ on the result of $\overline{n}$, the $\overline{B}$ values determined by the separate background measurements were used in the analysis of all ion spectra.

In the case of the hydrogen spectra (Fig.\,10\,left) there is a definite excess of measured events over the fitted line on the left hand side of the single ion peak \footnote{This effect was only observed in the case of the hydrogen spectra where the signal peak position is far from the background, whereas for the heavier ones the Poissonians overlap such that the effect can not be noticed.}. Such a background could be either produced by a relatively small number of ions that reached the crystal surface with a reduced energy or by backscattered ions which deposit only part of their energy in the crystal. In order to address this question, the process of the ions impinging onto the crystal surface was simulated using the SRIM2003 simulation package (\underline{S}topping and \underline{R}ange of \underline{I}ons in \underline{M}atter version \underline{2003})\cite{srim2003}. This package calculates the stopping power and the range of ions in matter using a full quantum mechanical treatment of ion-atom collisions. Here, only the effect of backscattered ions will be discussed.

Figure\,11 shows the simulated spectra of the energy deposition in the target (\ca) crystal from the backscattered ions for several projectiles. The total number of simulated ions was the same for all elements (10000 ions) and the numbers of backscattered ones are indicated in the figure. The energy spectra as well as the probabilities for the backscattering obtained from the simulation were implemented in the fit using the following procedure. An energy deposition with an energy lower than the nominal one will result in the production of a reduced number of photons obeying Poisson statistics within each energy bin. In order to consider energy losses due to the backscattering, the following fit function was used
\begin{footnotesize}
\begin{equation}
\hspace*{-10mm}
P_t(k)=P_{ion}(0)  P(k,\overline{B})+\sum_{i=1}^{N}P_{ion}(i) \sum_{l=0}^{k} P(l,\overline{B}) \int_0^{E_{ion}}P_S(k-l,\overline{n}\, i\frac{E_d}{E_{ion}}) P_{BS}(E_d) dE_d\,,
\end{equation}
\end{footnotesize}
where $P_{BS}(E_d)$ is the probability density that due to the backscattering losses the energy $E_d$ is deposited instead of the full energy of the ion $E_{ion}$. $P_{BS}(E_d)$ has been calculated with SRIM2003.
Fits including the effect of backscattering are shown in figure\,10 as solid lines. Adding the backscattering effect gave a better agreement with the measured data in the case of the lighter elements. Therefore, the backscattering of the ions can be only a partial explanation of the excess events observed on the left side of the single ion peak. No change was observed for the heavier ones where the backscattering effect is only in the permil range.

For the light ions like hydrogen another possibility for energy losses is conceivable. Ions hitting the collimator inner surface could be scattered with a shallow angle and therefore reach the crystal surface with a reduced energy. In order to address this issue, a test measurement was performed where a thin kapton tape was glued on top of the existing ion collimator hole ($\o$\,=\,1\,mm). A hole with a diameter of 0.7\,mm was made in the center to prevent the ions from hitting the surface of the long collimator hole. This way, the number of the events that can scatter off the teflon surface should be considerably reduced. The resulting spectra showed a lower relative intensity of the excess events, but most probably due to a non perfect alignment of the used collimator, the effect was still partially observed. This indicates that the nature of the small excess background, observed in the case of H ions, in the region below the single ion peak can be at least partially explained, but further investigation is needed for a full understanding. Nevertheless, these excess events are only a very small fraction of the total counts and have a negligible influence on the number of photoelectrons per single ion hit resulting from the fit.

In the case of heavier ions the fit was more sensitive to the number of Poisson distributions included in the fit due to the strong overlap of the distributions (see Fig.\,10). As previously discussed, via the consistent use of defocusing of the beam for all ions the probability for multiple ion events was reduced to a very low level and therefore only single and a small contribution of double ion events had to be taken into account for the determination of $\overline{n}$.
The introduction of a triple ion contribution into a fit of the W data leads to a strong correlation of the fit parameters resulting in unphysical intensity relations ($P_{ion}(3)>P_{ion}(2)$)\,(see Fig. 10). Nevertheless, the resulting $\overline{n}$ agree within statistical errors.

Technically it was not possible to obtain a beam with only 0 and 1 ion arriving due to the sharp dependence on the laser beam power and defocusing. This effect is illustrated in figure 9, where even for the smallest laser power and the weak focusing a certain contribution of two ion signals is apparent. Therefore the single ion model is excluded.

According to the previously discussed points, we decided to include the backscattering effect and to take into account only single and double ion contributions in the final fitting procedure.

The ratio of the number of registered photons from the ion spectra, $pe^-(ion)$, and the reference measurement, $pe^-(^{55}Fe)$, normalized (to 1\,keV), will give the quenching factor for that specific element:
\begin{equation}\label{eq:QFdef}
    QF(ion)\,=\,\frac{pe^-(^{55}Fe)}{pe^-(ion)}\,\cdot\,\frac{18\,keV}{5.9576\,keV}\,.
\end{equation}
where $5.9576\,keV$ is the weighted energy of K-lines from $^{55}Mn$ \cite{TOI1996}.

The linearity of the light output of the used sample was studied in a separate measurement with various gamma sources. The result is shown in  figure\,12. The observed good linearity justifies the reference measurement at 5.9\,keV.

\subsubsection{Quenching factor results}
Figure\,13 shows fitted spectra for oxygen, calcium and tungsten ions of 18~keV measured with a 40\,$\mu$s time window. Mean numbers of photons for single ion events obtained from the fits to spectra of all elements, as well as numerical values calculated for the quenching factors are given in table\,\ref{table:QFvalues}. A graphical representation of the dependence of the quenching factors on the atomic mass of the element, for 40\,$\mu$s time window, is given in figure\,14. Within the 30\,$\mu$s time window not all emitted light is collected (see Fig.\,7) whereas practically no light is emitted more than 40\,$\mu$s after the impact. This explains the slightly different mean values of $pe^-$ measured in the 30\,$\mu$s and 40\,$\mu$s time windows (see table\,\ref{table:QFvalues}). Practically identical values are obtained in the 40\,$\mu$s and 50\,$\mu$s time window measurements. Figure\,14 also includes results from neutron scattering experiment at room temperature \cite{coppi2005} and the values measured with a cryogenic detector at 7\,mK \cite{cresst2}. The presented data for oxygen and the neutron scattering result are in reasonable agreement. The low temperature values follow the systematic mass dependence of the room temperature results. This demonstrates the absence of a significant temperature dependence of the quenching factors in \ca.

The presented new method for the quenching factor measurements has several advantages compared to traditional neutron scattering experiments. The main advantage is that it is a simple “table-top” experiment. Measurements can be performed within a short period of time and a profuse choice of nuclei. A possible limitation may be due to surface effects causing a degradation of the light output for energy depositions close to the crystal surface. Therefore, the technique can be used in principle for any (solid) scintillation material, which does not suffer from a surface degradation of the scintillation efficiency. In the case of \ca crystals we could not observe any surface dependence. Strong arguments for this could be found in the comparison of the light yields from the $^{210}$Po alpha decays at the crystal surface and $^{210}$Po decays in the bulk of the crystal measured with cryogenic detectors (see \cite[Fig. 7 ]{cresst2}). The light yield is the same for both groups of events, showing, at least for alpha particles, that the quenching factor is the same for interior and surface events. This supplements the argument in \cite{meunier} that the absence of any splitting of the electron-photon band, where one has electrons mostly near the surface and photons mostly in the interior, demonstrates that bulk and surface events have the same quenching factor.

%

\section{Conclusion}
The quenching of the scintillation light from nuclear recoils is the key for an efficient background rejection in cold Dark Matter particle detectors that utilize scintillators. A new and powerful technique for the measurement of quenching factors was developed. It relies on the measurement of the number of photons produced when the scintillating crystal is bombarded by different ions. It uses a mass spectrometer to accelerate ions and measures, in single photon counting mode,  the light which they produce when impinging onto a crystal.

With the new technique presented here it was possible for the first time to determine the quenching factor of tungsten in \ca. The measurement of quenching factors of different recoiling nuclei allows their identification and offers practically a multiple target choice within one absorber crystal. This provides additional sensitivity by discrimination of neutron background (O\,-\,recoils) and possible \wimp events (W\,-\,recoils) \cite{cresst2}. In the case of a positive Dark Matter signal the change in the recoil energy spectra of different absorber nuclei gives a unique signature and a powerful tool for further identification and verification of \wimps properties.


\section*{Acknowledgment}
This work was partially supported by the DFG SFB 375 “Teilchen - Astrophysik” and the EU Network HPRN-CT-2002-00322 “Applied Cryodetectors”. The authors would like to thank Dr. Yorck Ramachers for bringing up the idea of using ions for quenching factor measurements. The mass spectrometer was funded by the Bundesministerium f\"{u}r Bildung und Forschung (BMBF) of the Federal Republic of Germany (FKZ 01 GG 9832).



\newpage
\begin{table}[h]
  \centering
  \caption{\small List of selected elements with associated target material.}\label{tab:elements}
 \begin{small}
  \begin{tabular}{|c||c||c||c|}
    \multicolumn{4}{c}{}\\
    \hline
    \xstrut Ion & Target material & Ion & Target material\\
    \hline \hline
    \xstrut H & Stainless steel &Y &  Y foil\\\hline
    \xstrut Be & Cu-Be foil &Ag &  Ag foil\\\hline
    \xstrut O &  CaO powder or \ca crystal& Sm &  Sm foil\\\hline
    \xstrut Si &  Stainless steel&W &  CuW foil and \ca crystal \\\hline
    \xstrut Ca &  CaO powder or \ca crystal&Au &  Au foil \\\hline
    \xstrut Cu &  Cu-Be foil&&\\
    \hline
  \end{tabular}
 \end{small}
\end{table}

\vskip4cm
\begin{table}[h]
\centering
\caption{\small Fitted values of the numbers of photoelectrons for the \fe peak. Three different time windows have been analyzed. The statistical errors (1$\sigma$) of the fits are quoted.}\label{tab:reference}
 \begin{tabular}{|c||c|}
   \multicolumn{2}{c}{}\\
   \hline
   \myVSpace.8cm
   \renewcommand\xstrut{\raisebox{-.4\myVSpace}{\rule{0mm}{\myVSpace}}}
   \xstrut \textbf{Time window}  & \textbf{Mean value}  \\
   \hline \hline
   \xstrut 30\,$\mu$s & 23.67\,$\pm$\,0.13\,pe$^-$  \\
   \hline
   \xstrut 40\,$\mu$s & 24.08\,$\pm$\,0.13\,pe$^-$   \\
   \hline
   \xstrut 50\,$\mu$s & 24.22\,$\pm$\,0.14\,pe$^-$   \\
   \hline
 \end{tabular}
\end{table}

\newpage
\begin{table}[h]
\centering
\caption{\small Comparison of fit parameters using different fitting procedures shown in figure\,10. The statistical errors (1$\sigma$) of the fits are quoted. $\chi ^2/dF$ is the chi squared per degrees of freedom and dF degrees of freedom. Discussion is given in the text.}
\scriptsize
 \begin{tabular}{|c||c|c|c|c|c|c|c|}
   \multicolumn{8}{c}{}\\
   \hline
Elem. & \textbf{Model}& $\chi^2$/dF&dF&\ $\overline{\textbf{n}}$&P$_{ion}$(1)&P$_{ion}$(2)&P$_{ion}$(3) \\        \hline\hline
  &      \begin{tabular}{c} Double ion events and \\ backscattering effect\\ \end{tabular} &57.2&73&\textbf{32.96$\pm$0.11}  &3968$\pm$64   &1117$\pm$37  & \\ \cline{2-8}
 \textbf{H} &      \begin{tabular}{c} Triple ion events and\\ backscattering effect\\ \end{tabular}&39.0&108&\textbf{33.00$\pm$0.11}&3963$\pm$64    &1149$\pm$37 &425$\pm$24 \\ \cline{2-8}
  &      \begin{tabular}{c} Double ion events and\\ NO backscattering effect\\ \end{tabular}&169223&71&\textbf{31.71$\pm$0.10}&3894$\pm$64&1092$\pm$37& \\ \hline\hline
 &      \begin{tabular}{c} Double ion events and\\ backscattering effect\\ \end{tabular}&6.4&3&\textbf{1.81$\pm$0.11}&1324$\pm$122&611$\pm$126& \\ \cline{2-8}
 \textbf{ W} &      \begin{tabular}{c} Triple ion events and\\ backscattering effect\\ \end{tabular}&2.0&4&\textbf{1.79$\pm$0.14}&1551$\pm$122&142$\pm$37&361$\pm$80 \\ \cline{2-8}
  &      \begin{tabular}{c} Double ion events and\\ NO backscattering effect\\ \end{tabular}&6.4&3&\textbf{1.81$\pm$0.11}&1324$\pm$122    &612$\pm$126  & \\ \hline
\end{tabular}\label{table:fitcomparison}
\end{table}

\newpage
\renewcommand{\baselinestretch}{1.}
\begin{table}[h]
  \centering
  \caption{\small Mean number of detected photoelectrons and calculated quenching factor values. The statistical errors (1$\sigma$) of the fits are quoted. $\chi ^2/dF$ is the chi squared per degrees of freedom, dF degrees of freedom and TW is the measurement time window.}\label{table:QFvalues}
  \scriptsize
  \begin{tabular}{|c||c||c||c|c|c|c|c||c|}
    \multicolumn{9}{c}{}\\
    \hline
\textbf{Element }& \textbf{TW}&$\overline{\textbf{n}}$&P$_{ion}$(1)&P$_{ion}$(2)&$\chi ^2$/dF&dF&P$_{ion}$(0)&\textbf{QF }\\ \hline\hline
        &\textbf{30$\mu$s}&\textbf{32.96 $\pm$ 0.11}&3968 $\pm$ 64&1117 $\pm$ 36&57.2&73&34079$\pm$185 &\textbf{2.17 $\pm$  0.02} \\ \cline{2-9}
\textbf{H}&\textbf{40$\mu$s}&\textbf{33.77 $\pm$ 0.11}&3965 $\pm$ 64&1110 $\pm$ 37&56.7&75&34062$\pm$185 &\textbf{2.15 $\pm$   0.02} \\ \cline{2-9}
        &\textbf{50$\mu$s}&\textbf{33.56 $\pm$ 0.11}&3933 $\pm$ 64&1105 $\pm$ 37&149.7&75&34135$\pm$185 &\textbf{2.18 $\pm$   0.02} \\ \hline\hline
        &\textbf{30$\mu$s}&\textbf{5.16 $\pm$ 0.07}&3082 $\pm$ 93&2436 $\pm$ 86&24.8&12&52011$\pm$230 &\textbf{13.9 $\pm$  0.2} \\ \cline{2-9}
\textbf{O}&\textbf{40$\mu$s}&\textbf{5.15 $\pm$ 0.07}&3062 $\pm$ 93&2467 $\pm$ 86&34.0&12&51960$\pm$230 &\textbf{14.1 $\pm$  0.2 }\\ \cline{2-9}
        &\textbf{50$\mu$s}&\textbf{4.93 $\pm$ 0.07}&2917 $\pm$ 97&2545 $\pm$ 90&35.98&11&51900$\pm$230 &\textbf{14.8 $\pm$  0.3} \\ \hline\hline
        &\textbf{30$\mu$s}&\textbf{3.36 $\pm$ 0.20}&373 $\pm$ 43&246 $\pm$ 41&5.48&7&16440$\pm$130 &\textbf{21.3 $\pm$  1.3 }\\ \cline{2-9}
\textbf{Si}&\textbf{40$\mu$s}&\textbf{3.39 $\pm$ 0.18}&376 $\pm$ 39&252 $\pm$ 37&9.00&7&16430$\pm$130 &\textbf{21.5 $\pm$  1.2}  \\\cline{2-9}
        &\textbf{50$\mu$s}&\textbf{3.31 $\pm$ 0.17}&385 $\pm$ 38&262 $\pm$ 36&9.99&7&16410$\pm$130 &\textbf{22.1 $\pm$  1.2} \\\hline\hline
        &\textbf{30$\mu$s}&  \textbf{ 2.57 $\pm$ 0.07 }&2780 $\pm$ 142&1616 $\pm$ 140&12.63&5&36750$\pm$195 &\textbf{27.8 $\pm$  0.8} \\\cline{2-9}
\textbf{Ca}&\textbf{40$\mu$s}&\textbf{ 2.68 $\pm$ 0.08}&2808 $\pm$ 141&1547 $\pm$ 139&14.17&5&36790$\pm$195 &\textbf{27.1 $\pm$  0.9} \\\cline{2-9}
        &\textbf{50$\mu$s}&\textbf{2.67 $\pm$ 0.08}& 2845 $\pm$ 141&1552 $\pm$ 139&16.05&5&36737$\pm$195 &\textbf{27.4 $\pm$  0.9 }\\\hline\hline
        &\textbf{30$\mu$s}&\textbf{2.37 $\pm$ 0.08}&1430 $\pm$ 85&873 $\pm$ 81&15.59&5&6309$\pm$82 &\textbf{30.1 $\pm$  1.1 }\\\cline{2-9}
\textbf{Cu}&\textbf{40$\mu$s}&\textbf{2.44 $\pm$ 0.08}&1411 $\pm$ 85&858 $\pm$ 81&14.84&5&6337$\pm$83 &\textbf{29.8 $\pm$  1.0} \\\cline{2-9}
        &\textbf{50$\mu$s}&\textbf{2.46 $\pm$ 0.08}&1406 $\pm$ 85&861 $\pm$ 82&14.68&5&6337$\pm$83 &\textbf{29.8 $\pm$  1.0} \\\hline\hline
        &\textbf{30$\mu$s}&\textbf{1.66 $\pm$ 0.03}&20086 $\pm$ 492&10878 $\pm$ 496&122.4&3&82670$\pm$310 &\textbf{43.1 $\pm$  0.9} \\\cline{2-9}
\textbf{Y}&\textbf{40$\mu$s}&\textbf{1.70 $\pm$ 0.03}&19893 $\pm$ 506&10769 $\pm$ 512&104.7&3&82920$\pm$310 &\textbf{42.8 $\pm$  0.8} \\\cline{2-9}
        &\textbf{50$\mu$s}&\textbf{1.70 $\pm$ 0.03}&19981 $\pm$ 510&10849 $\pm$ 515&105.8&3&82720$\pm$310 &\textbf{43.0 $\pm$  0.8} \\\hline\hline
        &\textbf{30$\mu$s}& \textbf{2.23 $\pm$ 0.10}&1244 $\pm$ 66&374 $\pm$ 63&98.11&4&129610$\pm$365 &\textbf{32.1 $\pm$  1.5} \\\cline{2-9}
\textbf{Ag}&\textbf{40$\mu$s}&\textbf{2.21 $\pm$ 0.10}&1352 $\pm$ 72&366 $\pm$ 70&164.2&4&129500$\pm$360 &\textbf{32.9 $\pm$  1.5} \\\cline{2-9}
        &\textbf{50$\mu$s}&\textbf{2.09 $\pm$ 0.08}&1505 $\pm$ 71&409 $\pm$ 64&162.7&4&129300$\pm$365 &\textbf{35.0 $\pm$  1.5} \\\hline\hline
        &\textbf{30$\mu$s}&\textbf{1.89 $\pm$ 0.04}&3771 $\pm$ 192&2197 $\pm$ 191&39.36&3&14795$\pm$130 &\textbf{37.9 $\pm$  0.9} \\\cline{2-9}
\textbf{Sm}&\textbf{40$\mu$s}&\textbf{1.92 $\pm$ 0.04}&3718 $\pm$ 200&2187 $\pm$ 199&32.76&3&14835$\pm$130 &\textbf{37.8 $\pm$  0.9} \\\cline{2-9}
        &\textbf{50$\mu$s}&\textbf{1.93 $\pm$ 0.04}&3720 $\pm$ 204&2202 $\pm$ 203&30.88&3&14808$\pm$130 &\textbf{37.9 $\pm$  0.9} \\\hline\hline
        &\textbf{30$\mu$s}&\textbf{1.72 $\pm$ 0.10}&1311 $\pm$ 121&687 $\pm$ 122&6.89&3&14220$\pm$125 &\textbf{41.6 $\pm$  2.4} \\\cline{2-9}
\textbf{W}&\textbf{40$\mu$s}&\textbf{1.81 $\pm$ 0.13}&1324 $\pm$ 136&611 $\pm$ 142&6.39&3&14280$\pm$125 &\textbf{40.1 $\pm$  2.7} \\\cline{2-9}
        &\textbf{50$\mu$s}&\textbf{1.83 $\pm$ 0.13}&1320 $\pm$ 137&605 $\pm$ 143&6.47&3&14290$\pm$125 &\textbf{40.0 $\pm$   2.7} \\\hline\hline
        &\textbf{30$\mu$s}&\textbf{1.76 $\pm$ 0.06}&5921 $\pm$ 260&2736 $\pm$ 268&42.5&3&121430$\pm$360 &\textbf{40.6 $\pm$  1.3} \\\cline{2-9}
\textbf{Au}&\textbf{40$\mu$s}&\textbf{1.87 $\pm$ 0.06}&6042 $\pm$ 254&2443 $\pm$ 268&56.11&3&121600$\pm$360 &\textbf{39.0 $\pm$  1.3} \\\cline{2-9}
        &\textbf{50$\mu$s}&\textbf{1.84 $\pm$ 0.06}&6313 $\pm$ 258&2471 $\pm$ 273&64.66&3&121280$\pm$360 &\textbf{39.7 $\pm$  1.3} \\\hline
\end{tabular}
\end{table}

\newpage \begin{figure} \centering \includegraphics[width=14cm]{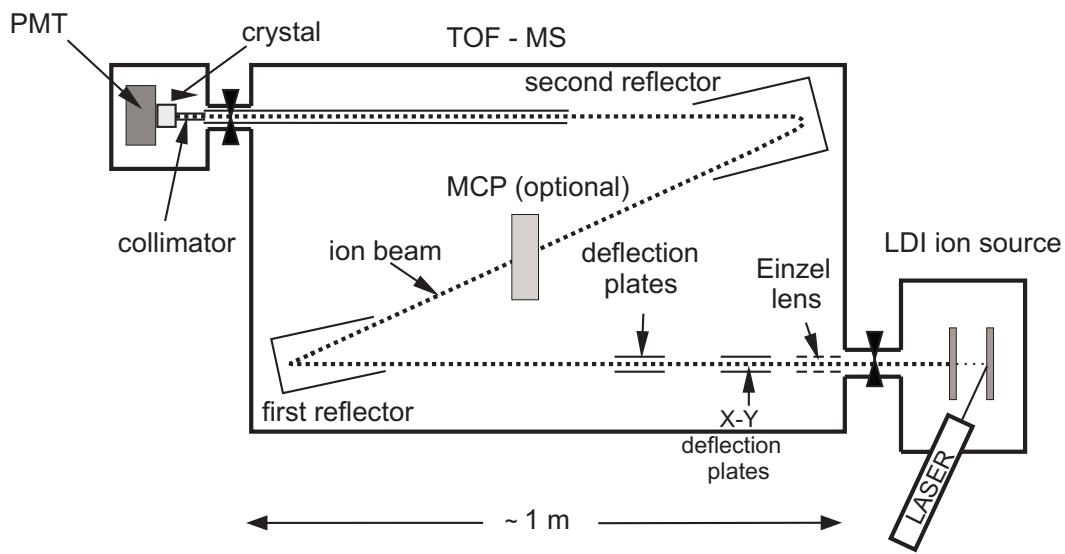} \\ \caption{Scheme of the experimental setup.} \end{figure}
\begin{figure} \centering \includegraphics[width=6cm]{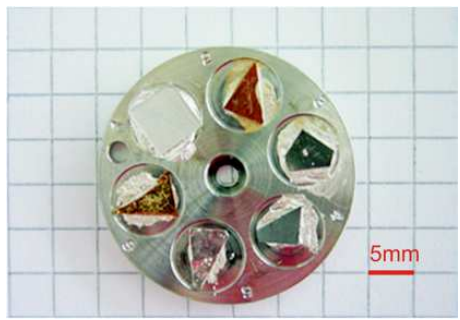} \\ \caption{Target holder with six different materials.} \end{figure}
\begin{figure} \centering \includegraphics[width=14cm]{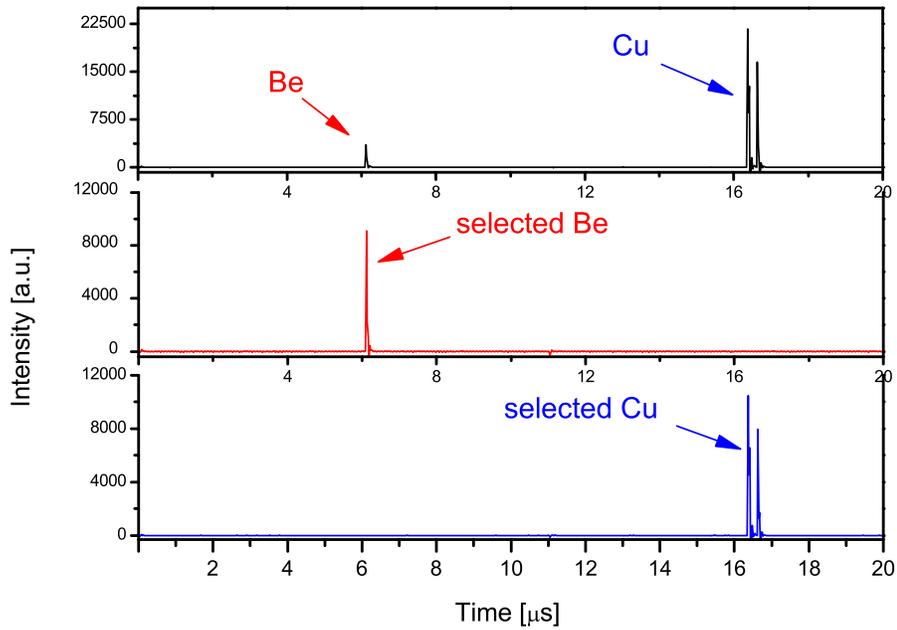} \\ \caption{Application of deflection plates for the selection of desired ions. Top: TOF spectrum of a CuBe target without deflection measured with the MCP. Applying the proper selection parameters only ions of beryllium (middle) or copper (bottom) are reaching the detector (and also the PM setup at the end of the TOF path).} \end{figure}
\begin{figure} \centering \includegraphics[width=8cm]{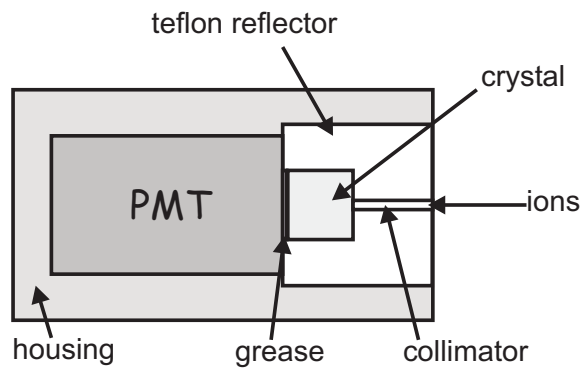} \\ \caption{Schematical view of the PM setup for photon counting (not to scale). The components are discussed in the text.} \end{figure}
\begin{figure} \centering \includegraphics[width=10cm]{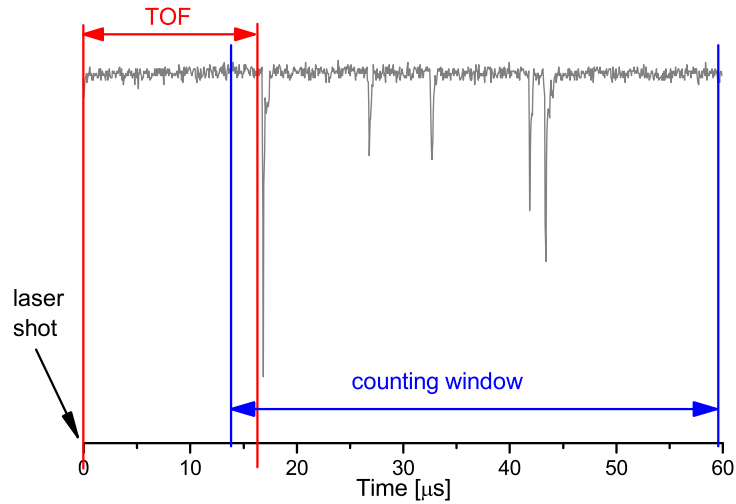} \\ \caption{Photomultiplier signal from one laser shot. The data acquisition is triggered by the laser. After the time-of-flight the ion hits the crystal. In the figure voltage pulses from single photons created by the ion are shown. Ions are counted within a fixed time window which opens shortly before the ion arrives after the known time of flight. The width of the counting window was kept constant for all measurements. } \end{figure}
\begin{figure} \centering \includegraphics[width=13cm]{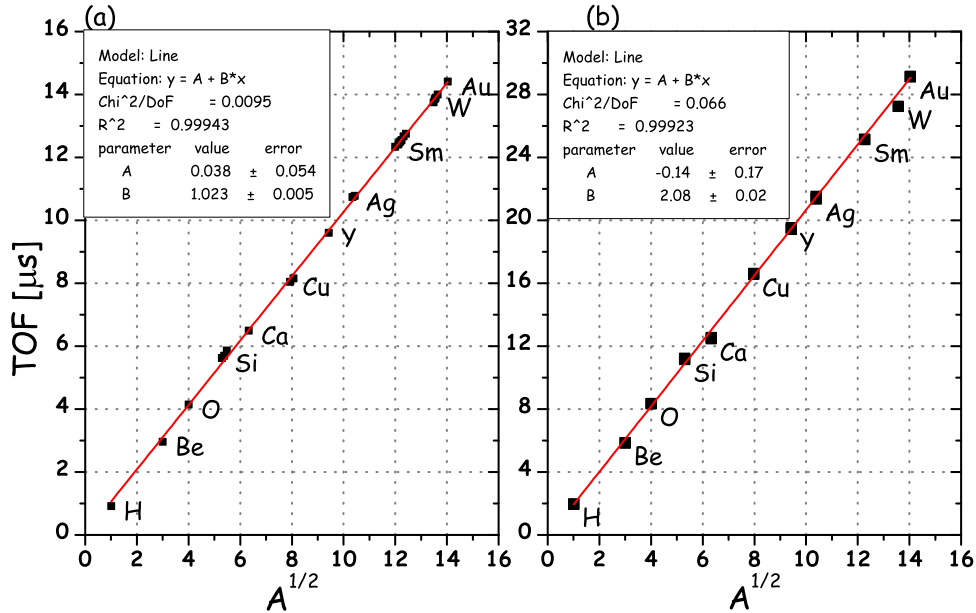} \\ \caption{Dependence of time-of-flight on $\sqrt{A}$ (where $A$ is a mass number) measured with the MCP placed at half the distance to the PM tube (a) and with a \ca crystal in the PM setup placed at the outlet of the mass spectrometer taking the onset of the light curve as the arrival time of the ion (b). The solid lines represent linear fits to the measurements with  parameters as given in the boxes. } \end{figure}
\begin{figure} \centering \includegraphics[width=14cm]{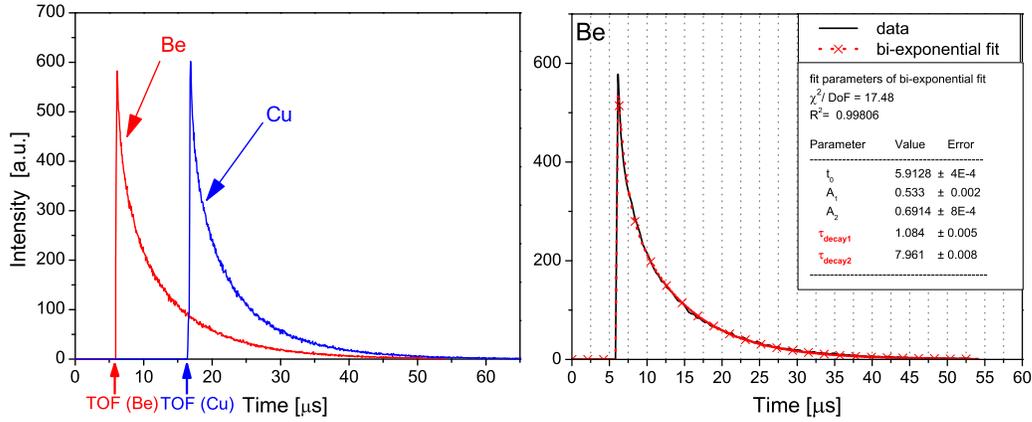} \\ \caption{Light curves for beryllium and copper (left) and bi-exponential fit of the light curve for beryllium (right). The fit parameters are indicated on the graph. The fast decay component is 1.08\,$\mu$s while the slow one is 7.96\,$\mu$s. } \end{figure}
\begin{figure} \centering \includegraphics[width=10cm]{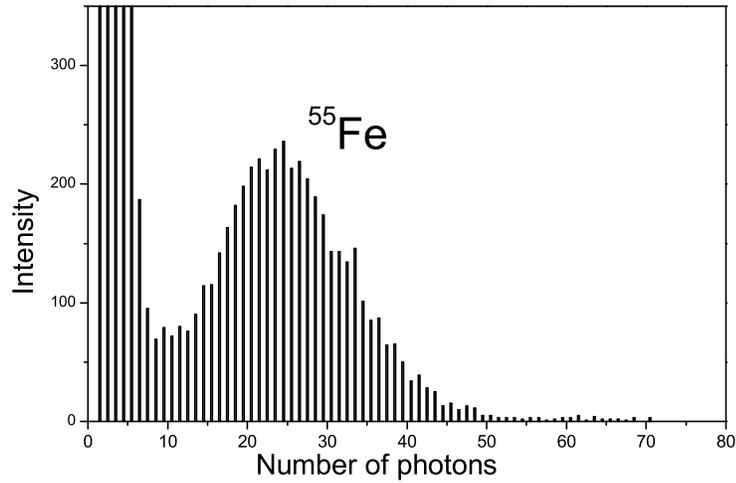} \\ \caption{Photon multiplicity spectrum recorded with a time window of 40\,$\mu$s while irradiating the \ca crystal with X-rays from \fe source. The corresponding fitted Poissonian mean value is 24.08$\pm$0.13\,pe$^-$ (4.04$\pm$0.02\,pe$^-$/keV).} \end{figure}
\begin{figure} \centering \includegraphics[width=14cm]{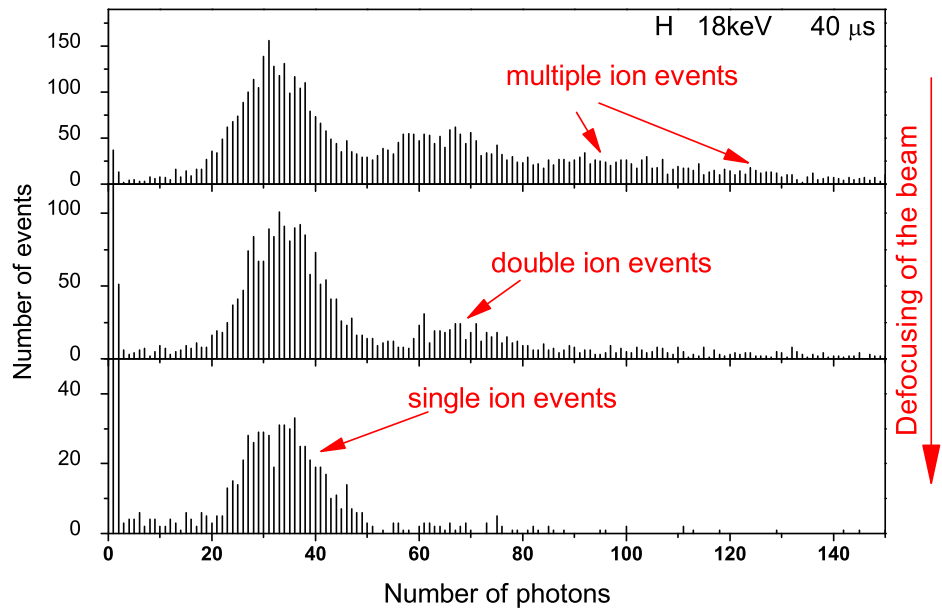} \\ \caption{Illustration of the effect of the beam defocusing on the shape of photon multiplicity spectra measured for hydrogen ions of 18\,keV. The total number of laser shots for the highest focusing is $\approx$\,8500 (top) while for the other two spectra this number is doubled. Defocusing reduces the probability of multiple ion events and significantly prolongs the measurement time. Technically it was not possible to obtain exclusively single ion production.} \end{figure}
\begin{figure} \centering \includegraphics[width=14cm]{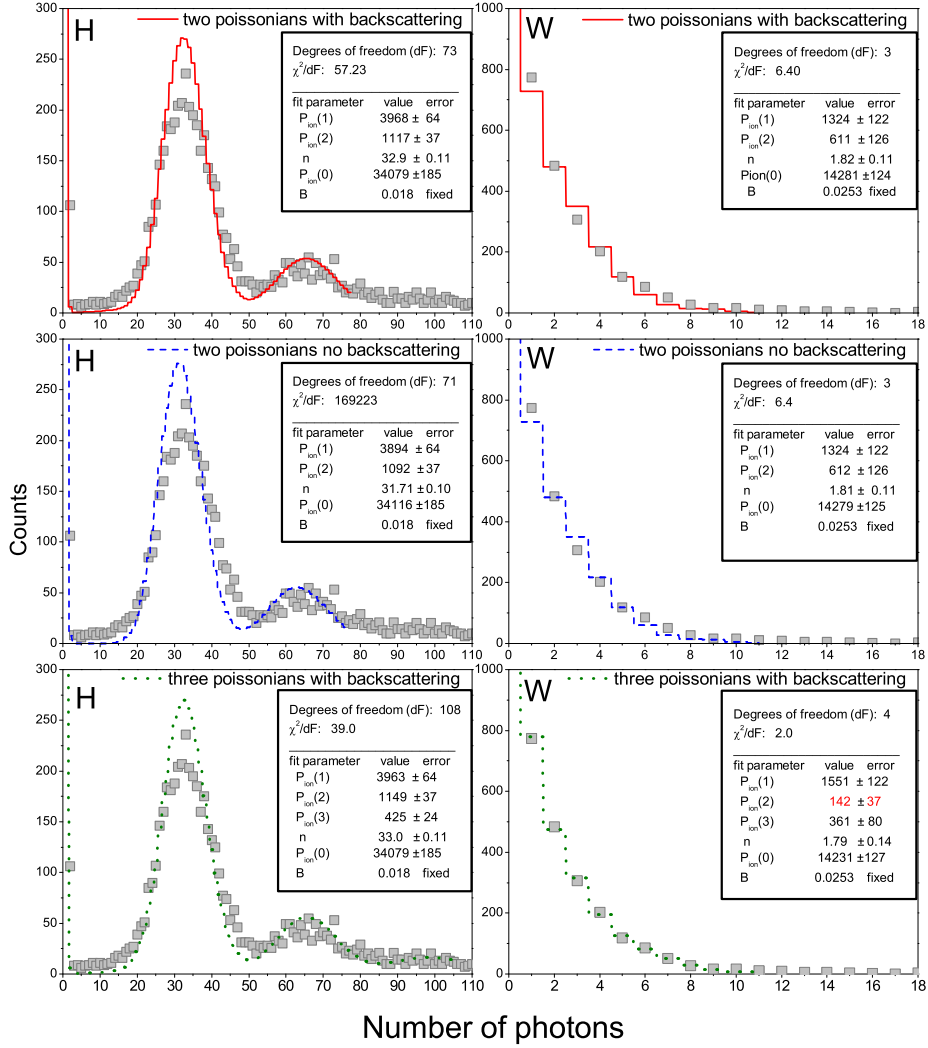} \\ \caption{Comparison of different fits to photon multiplicity spectra. Two examples are shown: hydrogen, 18\,keV (left) and tungsten, 18\,keV (right). In the case of hydrogen a higher laser intensity was chosen on purpose to produce a visible contribution of two and three ion events. For the tungsten case a lower laser intensity was chosen producing a small contribution of double ion and negligible contribution of triple ion events. The squares represent the measured data. The solid lines are likelihood fits including single and double ion events and backscattering of the ions. The dashed lines are fits without the backscattering effect while the dotted lines show the fits which include triple ion events and backscattering of the ions.  The data point with zero photon counts is far above the range shown. The fit parameters are given in boxes and table\,\ref{table:fitcomparison}. The discussion is given in the text.} \end{figure}
\begin{figure} \centering \includegraphics[width=12cm]{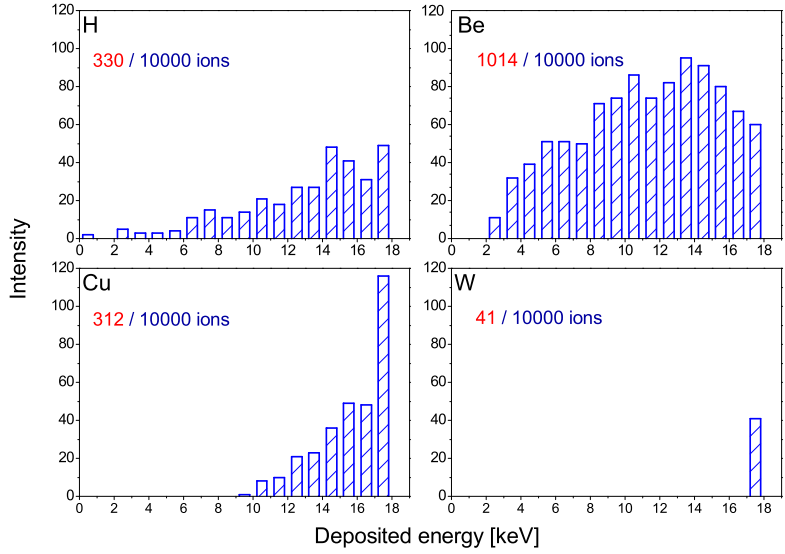} \\ \caption{Reduced energy deposition due to backscattered ions. A total of 10000 ions has been simulated with SRIM2003 package in each case. For each element the number of ion hits with a reduced energy deposition due to backscattering is indicated in the upper left corners of the graphs.} \end{figure}
\begin{figure} \centering \includegraphics[width=10cm]{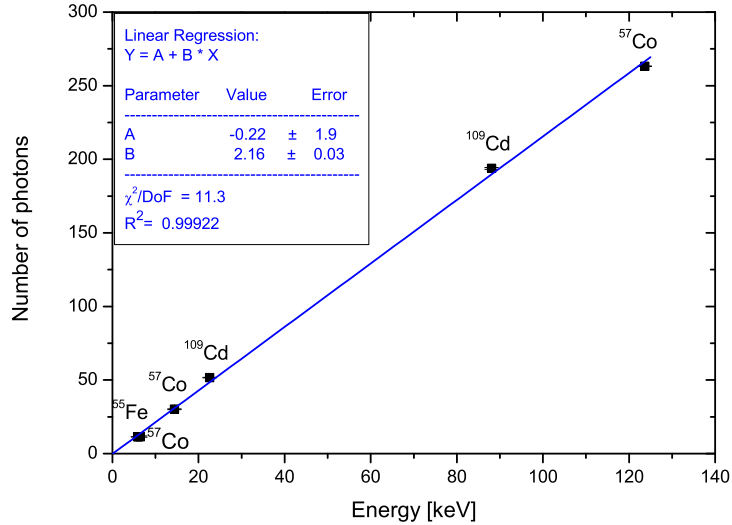} \\ \caption{Linearity of the light response of the \ca sample used for the measurements. The \ca crystal was irradiated with several \gam sources indicated on the graph and the photon counting was performed. The mean value for every detected line was determined as the mean of the Gauss distribution. The squares represent measured points while the line gives the linear fit to data. Fit parameters are indicated on the graph.} \end{figure}
\begin{figure} \centering \includegraphics[width=11cm]{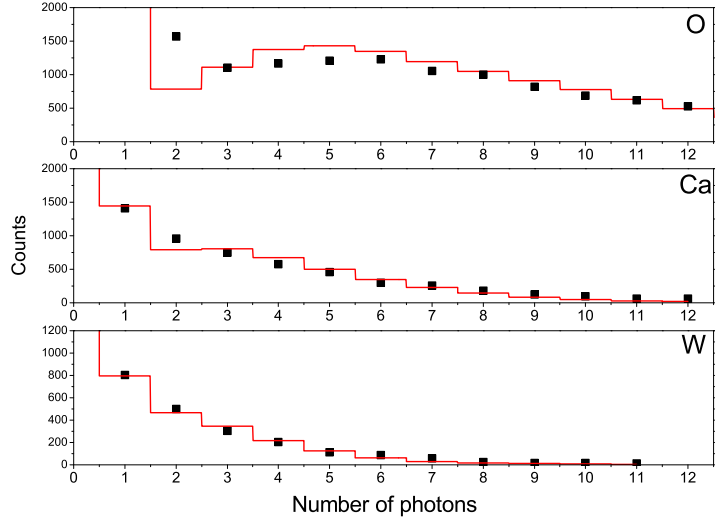} \\ \caption{Fit of measured spectra for oxygen (top), calcium (middle) and tungsten (bottom) ions. All spectra are measured with a 40\,$\mu$s time window. Corresponding fit values of the mean number of $pe^-$ per single ion hit are \textbf{O}:\,5.15\,$\pm$\,0.07\,pe$^-$, \textbf{Ca}:\,2.78\,$\pm$\,0.07\,pe$^-$ and \textbf{W}:\,1.81\,$\pm$\,0.13\,pe$^-$. } \end{figure}
\begin{figure} \centering  \includegraphics[width=11cm]{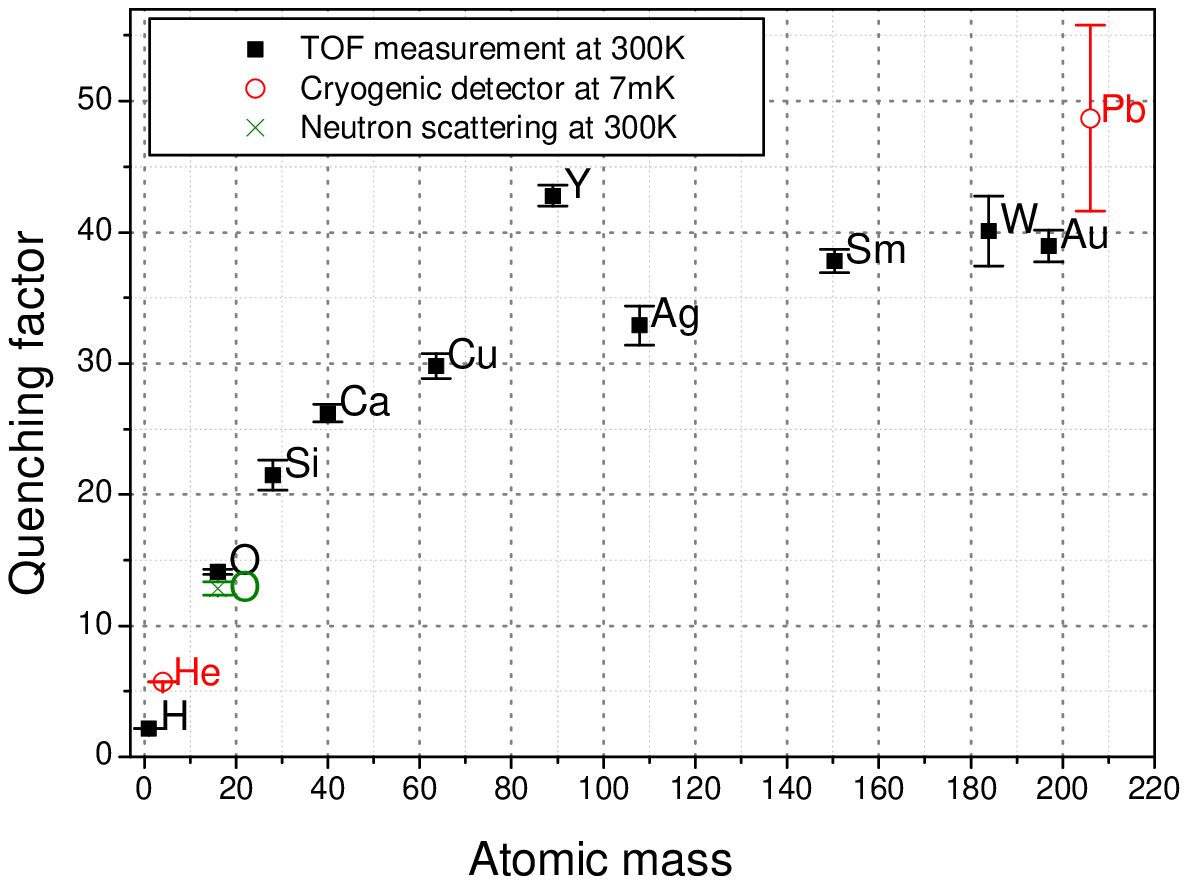} \\ \caption{Dependence of the quenching factor of \ca on the atomic mass of selected elements. The squares are the data of the presented work. The neutron scattering result (shown as $\times$) also measured at room temperature is from \cite{coppi2005}. The circles are from a measurement with a cryogenic detector at a temperature of 7\,mK. The Pb value, (48.7$\pm$7.1), for an energy of $\sim$\,100\,keV, is taken from \cite{cresst2}, while the He value (5.70$\pm$0.01) is derivated from the 2.3\,MeV $\alpha$-peak in \cite[Fig.7]{cresst2}. } \end{figure}


\newpage

\begin{thebibliography}{10}
\expandafter\ifx\csname url\endcsname\relax
  \def\url#1{\texttt{#1}}\fi
\expandafter\ifx\csname urlprefix\endcsname\relax\def\urlprefix{URL }\fi

\bibitem{cdms2004}
D.~S. Akerib, et~al., First Results from the Cryogenic Dark Matter Search in
  the Soudan Underground Laboratory, Physical Review Letters 93~(21) (2004)
  211301.

\bibitem{edelweiss2005}
V.~Sanglard, et~al., Final Results of the EDELWEISS-I Dark Matter Search with
  Cryogenic Heat-and-Ionization Ge Detectors, Physical Review D (Particles,
  Fields, Gravitation, and Cosmology) 71~(12)  122002.

\bibitem{cresst2}
G.~{Angloher}, et~al., J.~{Ninkovi\'{c}}, {Limits on WIMP Dark Matter using
  Scintillating CaWO$_4$ Cryogenic Detectors with Active Background
  Suppression}, Astropart. Phys.~(23) (2005) 325.

\bibitem{kamionkowski}
G.~Jungman, M.~Kamionkowski, K.~Griest, Supersymmetric Dark Matter, Phys. Rep.
  267 (1996) 195.

\bibitem{meunier}
P.~Meunier, et~al., Discrimination Between Nuclear Recoils and Electron Recoils
  by Simultaneous Detection of Phonons and Scintillation Light, Appl. Phys.
  Lett. 75~(9) (1999) 1335.

\bibitem{Lin65}
J.~Lindhard, Influence of Crystal Lattice on Motion of Energetic Charged
  Particles., Mat. Fys. Medd. Dan. Vid. Selsk. 34~(14).

\bibitem{lindhard}
J.~Lindhard, V.~Nielsen, M.~Scharff, Integral Equations Governing Radiation
  Effects, Mat. Fys. Medd. Dan. Vid. Selsk. 33~(10) (1963) 1--42.

\bibitem{Spooner1994}
N.~Spooner, et~al., The Scintillation Efficiency of Sodium and Iodine Recoils
  in a NaI(Tl) Detector for Dark Matter Searches, Phys. Lett. B 321 (1994)
  156--160.

\bibitem{tovey1998}
D.~Tovey, et~al., Measurements of Scintillation Efficiencies and Pulse-Shapes
  for Nuclear Recoils in NaI(Tl) and CaF$_2$(Eu) at Low Energies for Dark
  Matter Experiments, Phys. Lett. B 433 (1998) 150--155.

\bibitem{tovey}
D.~Tovey, et~al., A New Model Independent Method for Extracting Spin-Dependent
  (Cross Section) Limits from Dark Matter Searches, Physics Letters B 488
  (2000) 17--26.

\bibitem{horn}
D.~Horn, et~al., The Mass Dependance of CsI(Tl) Scintillation Response to Heavy
  Ions, Nucl. Instr. and Meth. A 320 (1992) 273--276.

\bibitem{Zde04}
Y.G. Zdesenko et~al., Scintillation properties and radioactive contamination of \ca crystal
  scintillators, Nucl. Instr. and Meth. A 538 (2005) 657--667.

\bibitem{proteom}
P.~Christ, R.~S., et~al., High Detection Sensitivity Achieved with Cryogenic
  Detectors in Combination with Matrix-Assisted Laser Desorption/Ionization
  Time-of-Flight Mass Spectrometry, Eu. J. Mass Spec.~(10) (2004) 469--476.

\bibitem{srim2003}
SRIMgroup, The Stopping and Range of Ions in Matter - SRIM2003,
  WEB:http://www.srim.org/SRIM/SRIM2003.htm.

\bibitem{TOI1996}
R.~Firestone, V.~Shirley, C.~Baglin, S.~Frank~Chu, J.~Zipkin, The Table of
  Isotopes, 8th Edition, John Wiley and Sons, Inc., 1996.

\bibitem{coppi2005}
C.~Coppi, et~al., Quenching Factor Measurement for CaWO$_4$ by Neutron
  Scattering, in: Proceedings of the LTD-11, 2005, to appear.


\end{thebibliography}
\end{document}